\documentclass[aip,twocolumn,showpacs,amsmath,amssymb,apl,reprint]{revtex4-1}
\usepackage{graphicx}% 
\usepackage{dcolumn}% Align table columns on decimal point
\usepackage{bm}% bold math
\usepackage{braket}
\begin{document}

\title{Attosecond electro-optic effect in zinc sulfide induced by a laser field}
\author{T. Otobe}
\affiliation{Kansai Photon Science Institute, National Institutes for Quantum and Radiological Science and Technology, 8-1-7, Umemidai,
 Kizugawa City, Kyoto 619-0215, Japan}

%=====================================================================================================================
\begin{abstract}
An ultrafast anisotropic electro-optic effect of the zinc sulfide crystal is predicted employing a numerical pump-probe simulation. 
The numerical results indicate that the time-dependence of the anisotropic response of ZnS
exhibits a phase shift with respect to the pump laser field.
The phase shift coincides with the time-resolved dynamical Franz--Keldysh effect, which is the modulation of the isotropic part of the dielectric function.
While the probe frequency dependence around the band gap is not intense, it becomes intense at higher photon energies of approximately 42~eV.
\end{abstract}
%\keywords{time-resolved dynamical Franz--Keldysh effect; time-dependent density functional theory; zinc sulfide; optical anisotropy, Pockels-like effect}
%\pacs{42.65.Re, 78.20.Bh}
\maketitle
%=====================================================================================================================
%\section{Introduction}
In the last two decades, advances in laser sciences and technologies have led to the availability of intense coherent light sources with different characteristics.
Ultra-short laser pulses can be as short as few tens of attoseconds, leading to the development of the new field of attosecond science
\cite{Hentschel01}. 
Intense laser pulses of mid-infrared (MIR) or terahertz (THz) frequencies 
have also recently become available \cite{hirori11, Chin01}. 
By employing these extreme sources of coherent light, 
investigating the optical response of materials in real-time with sub-optical cycle resolution is possible
\cite{Hentschel01,Hirori11-2,Schiffrin12,Schultze13,Schultze14,Novelli13}. 
 
The dielectric function $\varepsilon_{\alpha\beta}(\omega)$ is the most fundamental 
quantity characterizing the optical properties of matter. 
The dielectric function observed in an ultrafast pump-probe experiment
can be further considered as a probe time ($T_p$)-dependent function, $\varepsilon_{\alpha\beta}(T_p,\omega)$.
We determined the sub-cycle change in the optical properties, i.e., the time-resolved dynamical Franz--Keldysh effect (Tr-DFKE),
which corresponds to the response of the dressed states and quantum path interference of different dressed states 
\cite{Jauho96,Nordstrom98,otobe16, otobe16-2,otobe17}.
In particular, this ultrafast change exhibits an interesting phase shift that depends on the field amplitude and probe frequency. 
By utilizing this phenomenon, we can develop an ultrafast optical modulator or an ultrafast optical switch.

Recently, the Tr-DFKE was experimentally observed by a near-infrared (NIR)-pump extreme ultraviolet (EUV)-probe with attosecond time resolution for polycrystalline diamond \cite{Lucchini16,Schlaepfer18}.
A similar effect was also observed in an excitonic state in a GaAs quantum well by THz-pump NIR-probe spectroscopy \cite{Uchida16}.
However, because the signal by the Tr-DFKE is small, the high precision measurement or intense pump field is required.  

The diagonal parts represent $\varepsilon_{\alpha\alpha}$ the isotropic response, whereas the off-diagonal parts 
$\varepsilon_{\alpha \beta}$ ($\alpha \ne \beta$) represent the anisotropic response.
Because the anisotropic response can be detected as the polarization direction, it is sensitive to the change of signal. 
The magnetic field and electric field can induce anisotropic properties in isotropic materials. 
An ultrafast optical Faraday effect induced by the circularly laser field has been theoretically proposed \cite{Wismer17}
Under the electric field, some material shows an intense electro-optic effect, e.g., the Pockels effect and the Kerr effect.
The electro-optic effect is utilized to probe the waveform of the THz field, and ultrafast phenomena such as 
 laser-accelerated electron bunch\cite{Nishiura17}.

The Pockels effect induced by the modulation of the crystal structure on the picosecond timescale is strong and is thus frequently employed.
In contrast, the electro-optic effect on the atto- or femtosecond timescale is attributed to electron dynamics.
The electro-optic effect of materials under an intense laser field on the attosecond timescale may differ from the electro-optic effect at longer time-scales
because the isotropic part of dielectric function $\varepsilon_{\alpha\alpha}$ is modulated non-adiabatically.
In this study, we would like to demonstrate the ultrafast electro-optic effect in the attosecond 
time domain by employing time-dependent density functional theory (TDDFT).
We assume ZnS as the target material, which is a typical electro-optic material.

%\section{Computational method}
To derive time-dependent conductivity, we will revisit a simple model that we reported in our previous work \cite{otobe16}.
The probe's electric field is assumed to be weak enough that it can be represented by linear response theory. We denote the electric current caused by 
the probe field as $J_p(t)$, which is assumed to be parallel to the
direction of the probe's electric field. 
Its relationship to the time-domain 
conductivity $\sigma(t,t')$ is given as:
\begin{equation} 
J^p_{\alpha}(t) = \int_{-\infty}^t dt' \sigma_{\alpha\beta}(t,t') E^p_{\beta}(t'), 
\label{def_sigma} 
\end{equation} 
where $E_p(t')$ is the electric field of the probe pulse.
We note that the conductivity $\sigma(t,t')$ depends on both 
times $t$ and $t'$ rather than the just the time difference
$t-t'$, owing to the presence of the pump pulse.
If the probe laser duration is much shorter than the optical cycle of the pump laser and 
has peaks at time $t=T_p$, we can define the time-dependent conductivity, $\tilde{\sigma}_{\alpha\beta}(T_p,\omega)$, as:
\begin{equation}
\tilde{\sigma}_{\alpha,\beta}(T_p, \omega) =\frac{\int dt e^{i\omega t} G(t)J^p_{\alpha}(t)}{\int dt E^p_{\beta}(t)},
%\sum_n e^{in\Omega T_p} \tilde{\sigma}^{(n)}(\omega+n\Omega),
%\label{td_sigma}
\end{equation}
where $G(t)$ is the probe pulse window function, $G(t)=e^{-(t-T_p)^2/\tau^2}$.

We use the  real-time TDDFT program package SALMON \cite{salmon}. 
The details of the computational methods have been reported elsewhere \cite{Bertsch00,Otobe08,otobe16}.
We describe the electron dynamics in a unit 
cell of a crystalline solid under a spatially uniform time-varying electric field $E(t)$. 
Treating the field as a vector potential $\vec A(t)=-c\int^t dt' \vec E(t')$, 
the electron dynamics are described by the %following 
time-dependent Kohn--Sham (TDKS) equation \cite{Runge84}. 
We use a norm-conserving pseudopotential for the electron-ion potential \cite{TM91,Kleinman82}. 
For the exchange-correlation potential, we employ an adiabatic 
local density approximation (LDA)\cite{PZ81}. 
The cubic unit cell containing 4 zinc atoms and
4 sulfur atoms was discretized into Cartesian grids of $24^3$.
The $k$ space also descretized into $16^3$ grid points

In practice, we use the following electric fields. 
The pump field is of the form
$E_P(t) = -E_{0,P} f_P(t) \cos \Omega t $ whose
direction is along the (001) axis.
The envelope is $f_P(t)=\cos^2\left( \frac{\pi }{ 2 T_P}t\right)$ for $-T_P<t<T_P$ 
and $f_P(t)=0$ for $|t| \ge T_P$.
The probe field is of the form
$E^{p}(t) = E^{p}_0 \sin(\omega_p t) 
\exp\left(-(t-T_p)^2/2\eta^2\right)$, oriented in the [100]
direction. 
The field strength is set to 
$E_{0,p}=2.7 \times 10^{-3}$ MV/cm, which is small enough
to probe the linear response of the medium.

%\section{Numerical results}
%\subsection{Around the band gap}
\begin{figure}
\includegraphics[width=80mm]{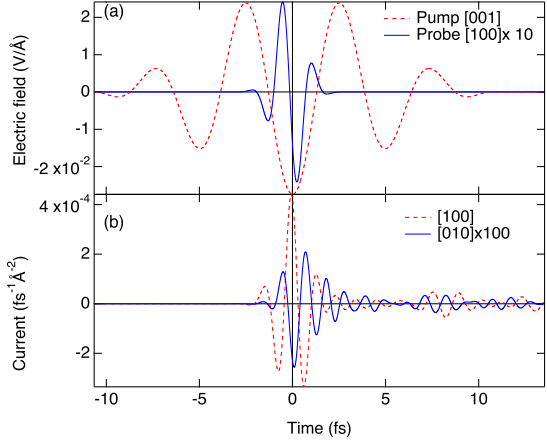}
\caption{\label{fig:Fig1} (a) Pump (red dashed line) and probe field (blue line) as a function of time.
The polarization of the pump pulse is parallel to the [001] (z-) direction and that of the probe pulse is parallel to [100] (x-) direction.
(b) Electronic current induced parallel (red dashed line) and orthogonal (blue solid line) to the probe pulse polarization.}
\end{figure}

\begin{figure}
\includegraphics[width=80mm]{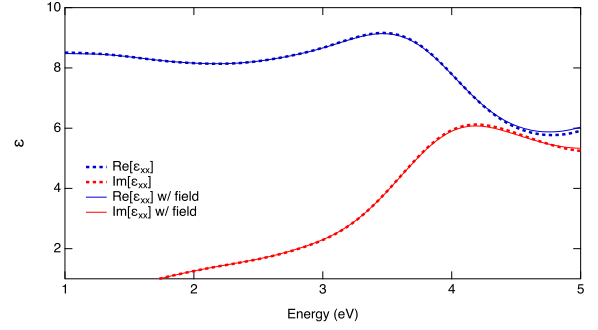}
\caption{\label{fig:Fig2}Dielectric function of ZnS with and without the pump field. The probe time ($T_p$) is 0~fs.}
\end{figure}

Typical calculation results are shown in Fig.~\ref{fig:Fig1}.
Figure \ref{fig:Fig1} (a) shows the electric field in the [001] (pump) and [100] (probe) directions.
The frequency of the pump field, $\Omega$, is 0.775 eV, and the pulse duration $T_P$ is 21.3~fs.
 The probe pulse duration ($\eta$) is set to 0.707~fs, and the center frequency is $\omega_p=2$~eV.
Figure \ref{fig:Fig1} (b) shows the induced current. 
The dashed red line and solid blue line represent the isotropic and anisotropic parts of the current, respectively. 
Figure \ref{fig:Fig2} shows the real and imaginary parts of $\varepsilon_{xx}$ with (dashed lines) and without (solid lines)
 the pump field calculated from the results shown in Fig. \ref{fig:Fig1}. 
 In this calculation, we use $\tau=3$~fs.
 
\begin{figure}
\includegraphics[width=80mm]{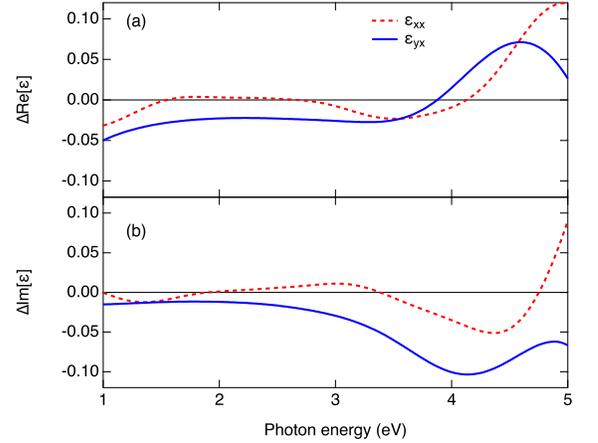}
\caption{\label{fig:Fig3} Modulation of the (a) diagonal and (b)off-diagonal part of the dielectric function.}
\end{figure}
Figure \ref{fig:Fig3} shows the difference between $\varepsilon$ with and without the pump fields.
The red lines present the diagonal part, $\Delta \varepsilon_{xx}(T_p,\omega)$, 
and the blue lines present the off-diagonal part, $\varepsilon_{yx}(T_p,\omega)$, which are induced by the pump laser field.
$\Delta \varepsilon_{xx}(T_p,\omega)$ is large around 4eV which corresponds to the intense absorption
above the band gap (red lines in Fig.~\ref{fig:Fig2}).
This result indicates that the Tr-DFKE and off-diagonal modulation in $\varepsilon$ are intense at 4eV, whereas 
diamond shows intense modulation around the optical band gap \cite{otobe16}.

\begin{figure}
\includegraphics[width=90mm]{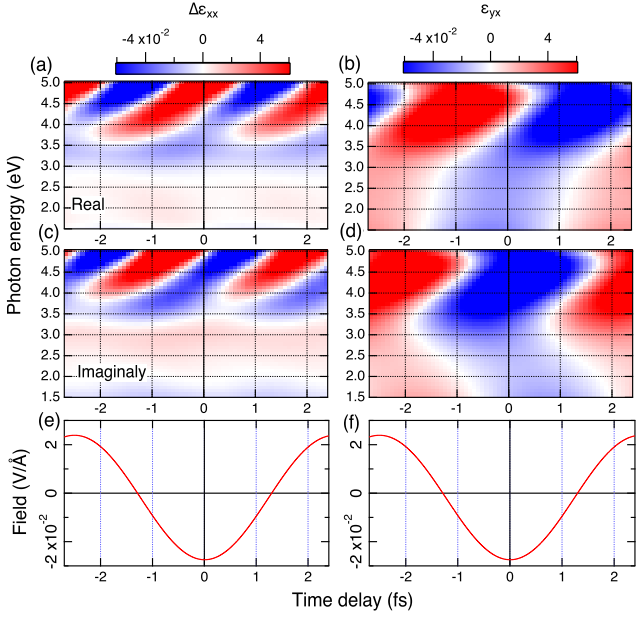}
\caption{\label{fig:Fig4}Time-resolved modulation of the dielectric function, $ \varepsilon_{\alpha\beta}(T_p,\omega)$. 
The peak intensity of the pump laser is $1\times 10^{10}$~W/cm$^2$ with frequency of $\Omega=0.775$~eV.
(a) (b) The real parts of $\Delta \varepsilon_{xx}(T_p,\omega)$ and $\varepsilon_{yx}(T_p,\omega)$. 
(c) (d) The imaginary parts of $\Delta \varepsilon_{xx}(T_p,\omega)$ and $\varepsilon_{yx}(T_p,\omega)$.
(e) (f) Pump field at the probe time delay.}
\end{figure}

The time-dependence of $\Delta \varepsilon(T_p,\omega)$ is shown in Fig.~\ref{fig:Fig4}.
ZnS shows intense modulation around 4 eV where intense photoabsorption occurs. 
The off-diagonal part $\varepsilon_{yx}$ shows odd-order response with respect to the pump electric field at each energy, 
whereas the $\Delta \varepsilon_{xx}$ has even-order due to the Tr-DFKE \cite{otobe16}.
This result indicates that the direction of the pump field switches the sign of $\varepsilon_{yx}$, 
which is qualitatively the same as odd-order nonlinear effects whose lowest order is the Pockels effect.

From the familiar formula for the Pockels effect for the cubic system, in the adiabatic limit, the anisotropic refractive index coincides with 
the pump laser field.
From previous works on the Tr-DFKE, $\Delta \varepsilon_{xx}(T_p)$ oscillates in even harmonic orders of $\Omega$,
 i.e. $e^{i2m\Omega T_p}$.
Then, $\varepsilon_{yx}$ shows the modulation with odd harmonics, $e^{i(2m\pm1)\Omega T_p}$.

Although the Pockels-like effect below 3~eV coincides with the phase of the pump laser field, 
the photon energy dependence in Fig.~\ref{fig:Fig4}(b) and (d) 
indicates a non-adiabatic response above 3.5~eV.
The phase shift of $\varepsilon_{yx}(T_p,\omega)$ with respect to $\omega$ is similar to that of $\Delta \varepsilon_{xx}(T_p,\omega)$.
The $\omega$ dependent phase shift in $\Delta \varepsilon_{xx}(T_p,\omega)$ corresponds to the relative phase of 
two Floquet states at $T_p$ \cite{otobe16,otobe16-2,Uchida16,otobe17}.
Therefore, the $\omega$-dependent phase indicates that the Pockels-like effect in the non-adiabatic regime 
is also the result of the relative phase between the Floquet states.

%\subsection{High photon energy region}
\begin{figure}
\includegraphics[width=80mm]{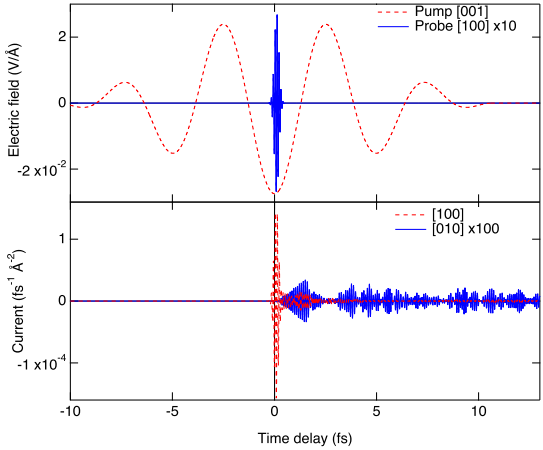}
\caption{\label{fig:Fig5} (Upper) Pump (red dashed line) and probe (blue line) fields as a function of time.
The polarization of the pump pulse is parallel to the [001] (z-) direction and the probe pulse is parallel to the [100] (x-) direction.
(Lower) Electronic current induced parallel (red line) and orthogonal (blue dotted line) to the probe pulse polarization.}
\end{figure}

\begin{figure}
\includegraphics[width=90mm]{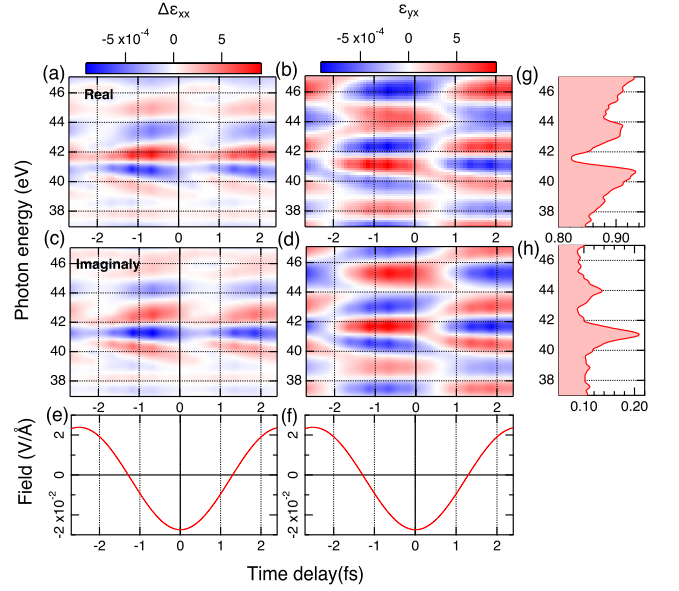}
\caption{\label{fig:Fig6}Time-resolved modulation of the dielectric function, $ \varepsilon_{\alpha\beta}(T_p,\omega)$. 
%The peak intensity of the pump laser is $1\times 10^{10}$~W/cm$^2$ at a frequency of $\Omega=0.775$~eV.
The center frequency of probe light ($\omega_p$) is 42~eV.
%(a) (b) The real part of the $\Delta \varepsilon_{xx}(T_p,\omega)$ and $\varepsilon_{yx}(T_p,\omega)$. 
%(c) (d) The imaginary part of $\Delta \varepsilon_{xx}(T_p,\omega)$ and $\varepsilon_{yx}(T_p,\omega)$.
%(e) (f) Pump field at the probe time delay.
%(g) (h)The real and imaginary part of $\varepsilon_{xx}$ w/o pump field.
The real and imaginary part of $\varepsilon_{xx}$ w/o pump field are shown in (g) and (h).}
\end{figure}

\begin{figure}
\includegraphics[width=90mm]{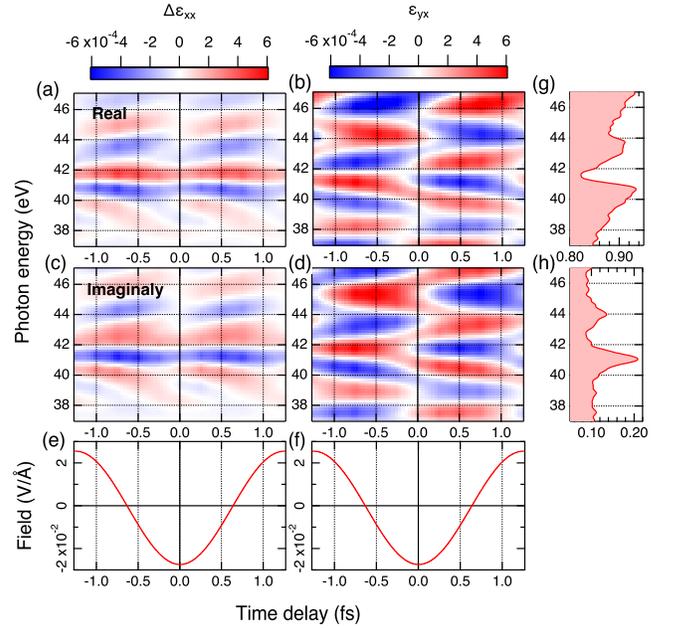}
\caption{\label{fig:Fig7}Time-resolved $\Delta \varepsilon_{\alpha\beta}(T_p,\omega)$. 
The peak intensity of the pump laser is $1\times 10^{10}$~W/cm$^2$ with  frequency of $\Omega=1.6$~eV.
}
\end{figure}

In the previous calculations, we showed the non-adiabatic sub-cycle response in the off-diagonal dielectric function in the proximity of the band gap.
However, sub-fs pulse in the VUV region have not been utilized. 
As the next step, we would like to show the sub-cycle response of $\varepsilon_{yx}(T_p,\omega)$ around the photon energy of 42 eV \cite{Lucchini16}.
In the case of the lower energy probe, interaction between a few bands may be dominant.
In particular, around the band gap, $\rm{Re}[\varepsilon_{yx}(T_p,\omega)]$ shows the usual Pockels-like effect adiabatic response. 
On the other hand, in the higher photon energy region, a response complicated by the contribution of many bands is expected.

Figure~\ref{fig:Fig5} shows (a) the pump and probe fields and (b) the induced current in the [100] and [010] directions.
The pulse duration $\eta$ is set to $\eta= 0.11$ fs.  
The anisotropic current (the current in the [010] direction) shows a slow increase compared to the case of the low frequency (Fig.~\ref{fig:Fig1} (b)).
This behavior indicates that the rotation of the polarization dose not only depends on the temporal anisotropy, but also the anisotropic current 
under the following pump pulse field. 
We set $\tau=2$~fs to include the peak of the anisotropic current around 2~fs in Fig.~\ref{fig:Fig5} (b).

The $\varepsilon(T_p,\omega)$ is shown in Fig.~\ref{fig:Fig6}. 
The diagonal part, $\Delta\varepsilon_{xx}(T_p,\omega)$, indicates Tr-DFKE occurs around 41 eV, which corresponds to the peak of the 
absorption without the pump field (Fig.~\ref{fig:Fig6} (g) and (h)).
 Although the off-diagonal part, $\varepsilon_{yx}(T_p,\omega)$ (Fig.~\ref{fig:Fig6} (b) and (d)), 
 indicates odd-harmonic oscillation with respect to the 
pump laser field as we expected, the direction of the light rotation is strongly dependent on the photon energy. 
The $\varepsilon_{yx}(T_p,\omega)$ also indicates intense phase shift with respect to the pump field, which coincides with 
the phase of the Tr-DFKE ($\Delta\varepsilon_{xx}(T_p,\omega)$).
These photon energy and time-delay dependencies do not appear in the low-frequency probe.

The Pockels-like response is expected to be sensitive to the pump laser frequency ($\Omega$), 
because the photon energy and delay-time dependence of $\varepsilon_{yx}(T_p,\omega)$ appear to be affected by that of $\varepsilon_{xx}(T_p,\omega)$.
Figure~\ref{fig:Fig7} shows the case of $\Omega=1.6$~eV.
Since the frequency is increased, the dynamical effect in $\varepsilon(T_p,\omega)$ should be enhanced compare to that of Fig.~\ref{fig:Fig6}.

The $\varepsilon_{yx}(T_p,\omega)$ and $\Delta\varepsilon_{xx}(T_p,\omega)$  show the maximum when $E_P=0$~V/\AA.
This behavior is similar to the Tr-DFKE with a weak pump laser field in diamond \cite{otobe16}.
With respect to the photon energy dependence, $\Delta\varepsilon_{xx}(T_p,\omega)$ in Fig.~\ref{fig:Fig7} shows a different dependence from Fig.~\ref{fig:Fig6},
because the energies of the Floquet states are different. 
On the contrary, the $\varepsilon_{yx}(T_p,\omega)$ shows almost the same photon energy dependence as that of Fig.~\ref{fig:Fig6}.

These results for the high-frequency probe case indicate that the anisotropic response including many electronic bands shows 
more complicated behavior than the Tr-DFKE and the low-frequency case.

%\section{Summary}
In this study, we demonstrated the ultrafast Pockels-like response in ZnS on the attosecond timescale using time-dependent density functional theory.
Our results demonstrate the usual adiabatic anisotropic response around the band gap.
Conversely, for the higher probe frequency, the $\varepsilon_{yx}$ shows significant dependence on the probe frequency and time.
In particular, the time dependence  coincides with the phase of the Tr-DFKE.
Because detecting the anisotropic response is sensitive to the modulation, it may be a good candidate for the new ultrafast optical switching device.

\section*{Acknowledgement}
This work was supported by JST-CREST under grant number JP-MJCR16N5, and by JSPS KAKENHI, Japan Grant Numbers 
15H03674 and JP17H03525. 
Numerical calculations were performed on the supercomputer SGI ICE X at 
the Japan Atomic Energy Agency.

\bibliography{DFKE_Pock.bib}

\end{document}